\documentclass{egpubl}
\usepackage{eurovis2026}

\EuroVisShort  

\usepackage[T1]{fontenc}
\usepackage{dfadobe}

\usepackage{cite}  
\BibtexOrBiblatex
\electronicVersion
\PrintedOrElectronic
\ifpdf \usepackage[pdftex]{graphicx} \pdfcompresslevel=9
\else \usepackage[dvips]{graphicx} \fi

\usepackage{egweblnk}

\usepackage{amsmath,amsfonts, subcaption}
\usepackage{xcolor}
\usepackage{comment}

\title[Summarizing Time-Varying DIC Strain Fields]%
{Summarizing Time-Varying Digital Image Correlation Strain Fields Using Sankey Diagrams}

\author[Victor Persson, Christofer Boo, Mohit Sharma \& Ingrid Hotz]
{\parbox{\textwidth}{\centering Victor Persson$^{1,2}$, Christofer Boo$^{2}$, Mohit Sharma$^{1}$ 
        and Ingrid Hotz$^{1}$ }
        \\
{\parbox{\textwidth}{\centering $^1$Link\"oping University, Sweden \\
$^2$Image Systems Motion Analysis}}} 

\begin{document}


\maketitle
\begin{abstract}
   Digital Image Correlation (DIC) enables dense, time-resolved measurement of surface strain in deforming materials, providing insight into strain localization and failure mechanisms. However, the resulting strain fields are typically explored frame-by-frame through spatial visualizations, making global temporal patterns difficult to discern. We present a visual summarization approach that represents the evolution of high-strain regions as a single Sankey diagram constructed from superlevel sets of the von Mises equivalent strain field. By tracking connected components over time via spatial overlap, the diagram encodes the birth, persistence, merging, and disappearance of strain concentrations. Applied to four tensile test datasets with varying notch geometries, the approach compactly captures differences in deformation regimes and qualitative precursors to failure, complementing traditional spatial strain visualizations with a global temporal overview.

\printccsdesc   
\end{abstract}  
\section{Introduction}

Understanding how materials deform and fail under load is a fundamental problem in experimental mechanics. While theoretical material limits provide important bounds, real-world failure is strongly influenced by geometry: features such as cutouts and notches act as stress concentrators, causing strain to localize and determine where and how rupture occurs \cite{MechanicalPhysicsBook}. Capturing and interpreting this behavior is critical for safe and reliable structural design.
Digital Image Correlation (DIC) is a standard technique for studying material deformation, enabling contactless measurement of dense surface displacement and strain tensor fields from image sequences. DIC produces high-resolution spatio-temporal datasets that reveal fine-grained strain patterns during mechanical tests~\cite{ImageCorrelationBook}. Modern DIC software systems also provide some visualization tools, such as color maps of derived scalars (e.g., von Mises strain), tensor glyphs (e.g., eigen-ellipses), and line-based methods like tensor line integral convolution (LIC)~\cite{Kratz2013,Hergel2021}.

This project was conducted in close collaboration with an industrial partner specializing in DIC image analysis software, providing access to real-world datasets and engineering workflows. Through this partnership, we identified limitations in current frame-based analysis practices, which require analysts to manually inspect hundreds of time steps to determine when high-strain regions appear, how long they persist, and how they interact prior to failure. As temporal resolution and dataset size increase, this process becomes inefficient and error-prone. These challenges motivated the methods proposed in this paper.
We propose a compact, topology-inspired visual summary of DIC strain data using Sankey diagrams. We interpret strain evolution as a time-varying scalar field and apply topological feature tracking for visualization. Specifically, we track connected regions of high von Mises strain, whose births, merges, splits, and disappearances encode meaningful deformation behavior. These events are summarized using Sankey diagrams to provide a concise representation of strain evolution over time. The approach is evaluated on four experimental datasets. This work makes the following contributions:
\begin{itemize}
\item A Sankey-based summary of time-varying strain fields that encodes the evolution of connected von Mises strain superlevel sets in a single diagram.
\item A lightweight, topology-inspired region-tracking method that captures region birth, persistence, merging, and splitting without explicit critical-point computation.
\item An evaluation on four tensile test datasets with varying notch geometries, demonstrating that the summaries distinguish deformation regimes and reveal failure precursors.
\end{itemize}


\section{Background}
This section outlines the application context, data acquisition process, and concepts underlying our visualization approach.

\begin{figure*}[!htbp]
    \centering
    \includegraphics[width=1\linewidth]{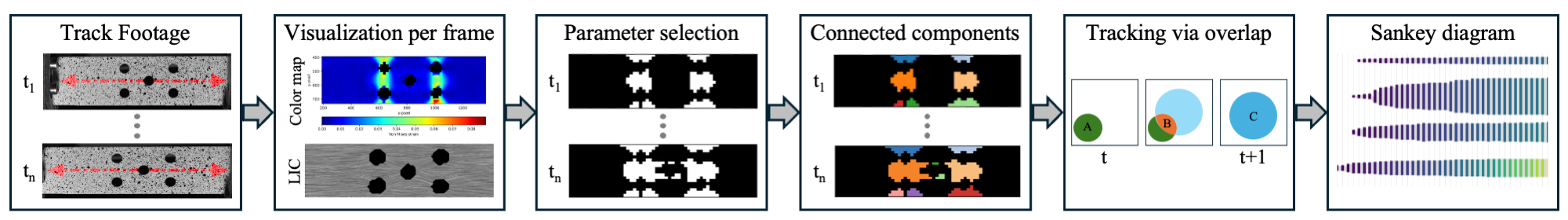}
    \caption{Visualization pipeline. Time-resolved DIC strain fields are processed per frame to compute the von Mises equivalent strain. LIC and color mapping provide an overview. Superlevel-set and persistence thresholds identify regions of interest, whose connected components are linked across timesteps by spatial overlap and summarized in a Sankey diagram. 
    }
    \label{fig:Pipeline}
\end{figure*}

\subsection{Tensile Testing and Strain Localization}
Tensile testing characterizes material response under uniaxial loading. A specimen of known geometry is subjected to increasing tensile force until elastic deformation transitions to plastic deformation and ultimately failure. While elastic deformation is reversible, plastic deformation causes permanent damage and is often accompanied by strain localization and necking.
Geometric irregularities such as notches act as stress concentrators, localizing strain and often governing the onset of yielding and fracture. Understanding how these strain concentrations evolve is central to predicting material failure~\cite{MechanicalPhysicsBook}. 

\subsection{Digital Image Correlation}
Digital Image Correlation (DIC)~\cite{DICpracticesGuide} is an optical, non-contact technique for measuring dense displacement and strain fields from image sequences. It tracks small subsets of pixels from a reference image across frames by minimizing intensity differences under a deformation model. Applying this over a dense grid produces time-resolved displacement fields over the specimen surface~\cite{ImageCorrelationBook}.

Strain tensors $\mathbf{F}=\mathbf{I} + \nabla \mathbf{u}$ are derived from these displacement fields $\mathbf{u}$ via spatial differentiation over a local window, $I$ is the Identity tensor. For the large deformations in tensile testing, finite-strain measures are required. Here, we use the Green–Lagrange strain tensor $\mathbf{E}=\tfrac{1}{2}\big(\mathbf{F}^{\top}\mathbf{F} - \mathbf{I}\big)$, defined relative to the undeformed configuration and invariant to rigid-body rotations~\cite{Holzapfel2000}. 

\begin{figure}[!htbp]
\centering
  \includegraphics[width=.95\linewidth]{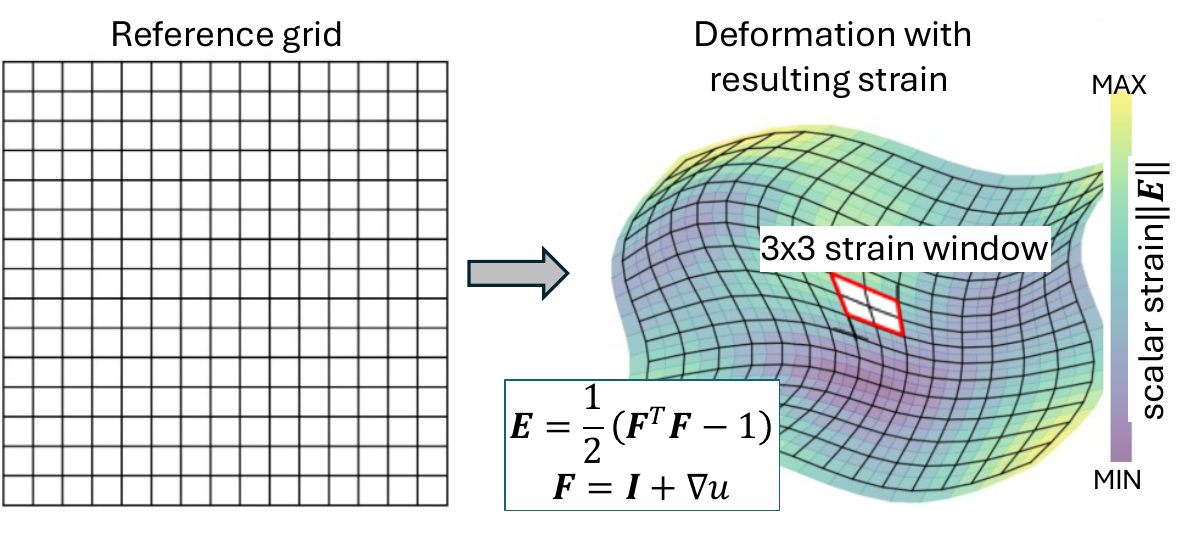}
  \caption{Calculating point-wise strain from a deformation field
}
  \label{fig:DIC to strain}
\end{figure}

\subsection{Strain Measures}
Strain is a symmetric tensor describing local stretching and shearing. Its eigenvalues and eigenvectors define the principal strain magnitudes and directions, which are commonly visualized using tensor glyphs or line-based methods.
To reduce tensorial strain to a single scalar field, the von Mises equivalent strain $\epsilon_{VM}$ is commonly used, expressed in terms of differences between the principal strains. This scalar captures distortional deformation while ignoring uniform volumetric changes, making it well-suited for uniaxial tensile tests. In this work, von Mises strain is the primary field used to identify and track regions of high stress.

\subsection{Superlevel Sets and Topological Evolution}
From a scalar-field perspective, high-strain regions correspond to connected components of superlevel sets of the von Mises strain field. Thresholding the field at a chosen value produces spatially coherent regions representing localized deformation, with connectivity typically defined using 8-neighbors to capture diagonally adjacent structures.
As loading increases, these components undergo topological changes: they appear, grow, merge, split, persist, or vanish~\cite{Yan2021b}. Their temporal evolution encodes key information about strain localization, necking, and the onset of failure. 

\subsection{Visualization and Temporal Aggregation}
Standard frame-based visualizations, such as scalar color maps, tensor glyphs, and LIC of principal strain directions (Fig~\ref{fig:LIC}), make temporal strain evolution difficult to interpret~\cite{Kratz2014}. Sankey diagrams offer a flow-based paradigm for summarizing how quantities change across discrete stages: nodes represent entities at each stage, and links encode persistence, splitting, or merging~\cite{Schmidt2008}. This makes them well-suited for tracking connected regions in time-varying scalar fields.
In this work, Sankey diagrams are used to aggregate and visualize the evolution of the von Mises strain field, providing a compact summary of strain evolution.


\section{Method}
This section describes the individual components of our methods summarized in the pipeline figure~\ref{fig:Pipeline}.

\subsection{Data Acquisition and Preprocessing}
The input data consist of time-resolved strain fields of INCONEL 718 tensile tests, derived from DIC displacement measurements performed using TrackEye Motion Analysis. From this the von Mises equivalent strain $\epsilon_{VM}$ is computed per frame. 
Spatial differentiation amplifies high-frequency noise, resulting in artifacts as subset jitter and spurious peaks near boundaries. To suppress these effects, each frame is filtered using a morphological $h$-maxima transform, which removes shallow local maxima while preserving physically meaningful structures.  
Invalid samples are replaced with the lowest valid values prior to filtering. Per-frame superlevel sets are then extracted by thresholding the filtered $\epsilon_{VM}$ field, and connected components are identified using 8-connectivity.

\subsection{Principal Strain Field Visualization}
\begin{figure}[!hb]
    \centering
    \includegraphics[width=0.9\linewidth]{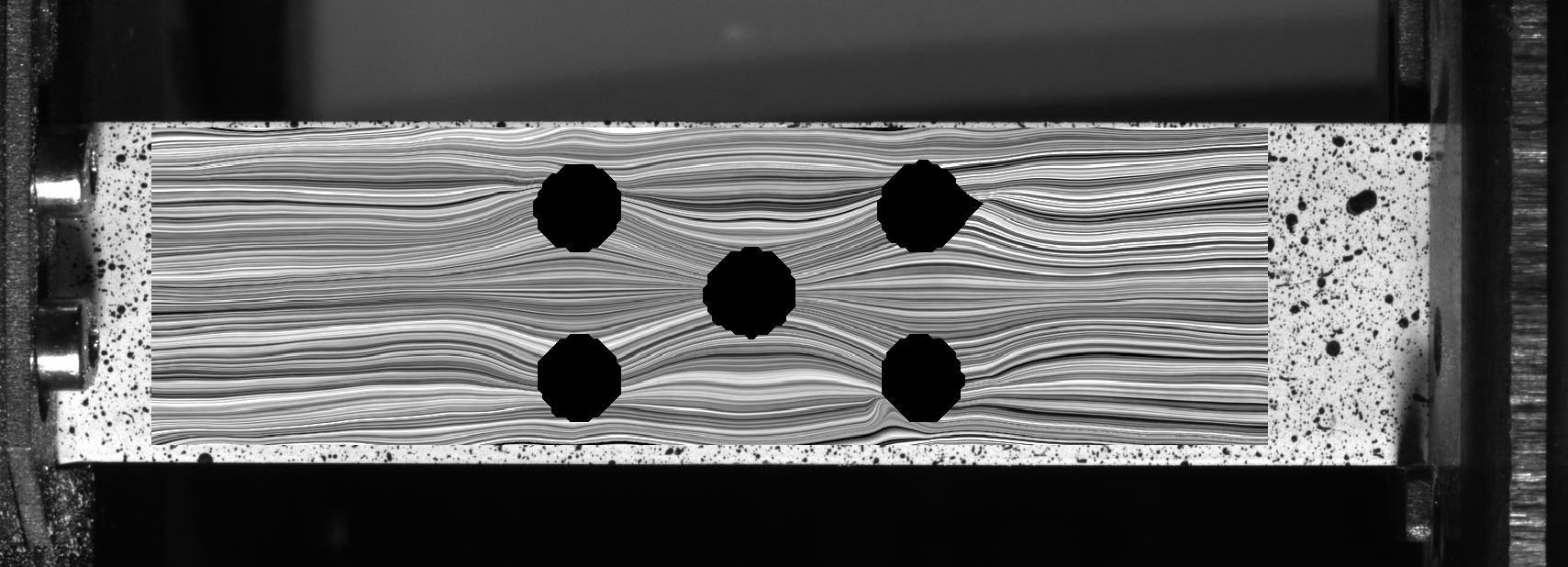}
    \includegraphics[width=0.9\linewidth]{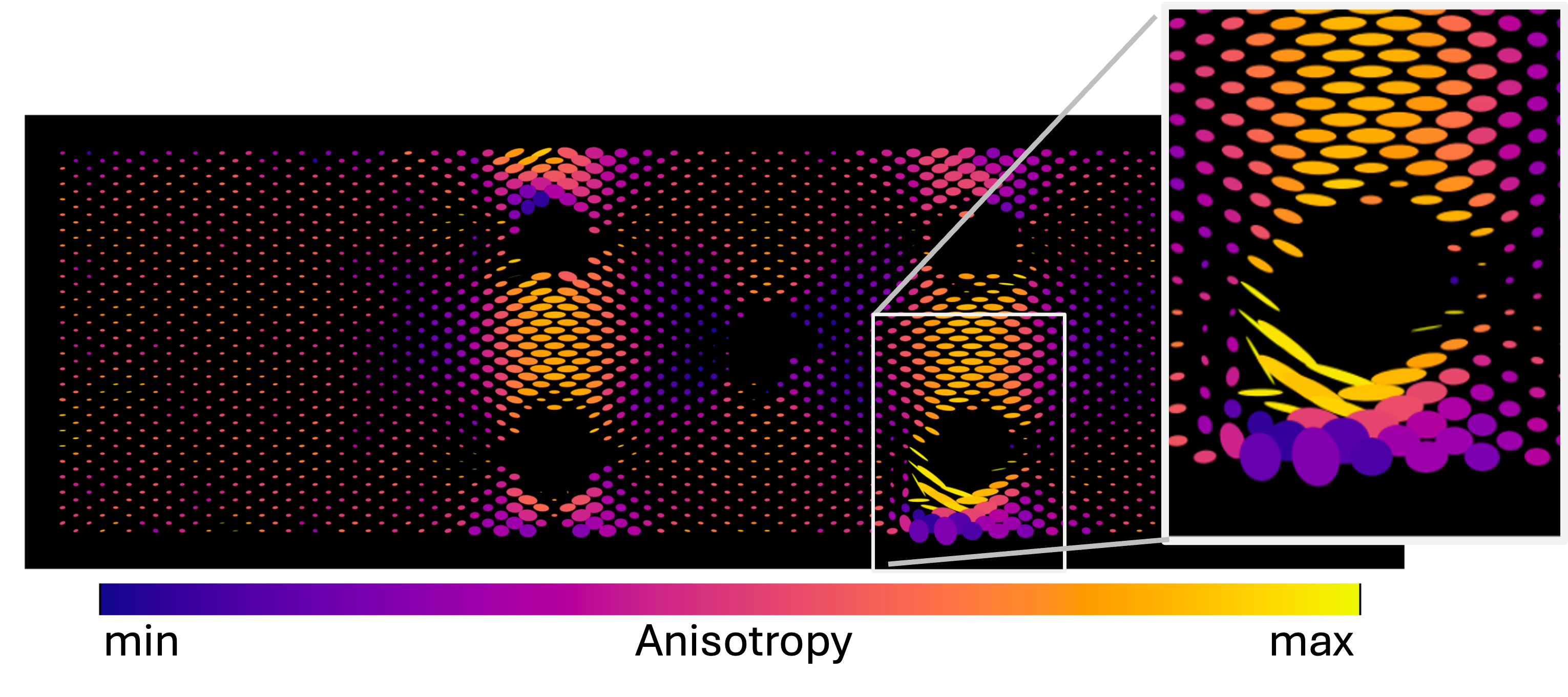}
    \caption{Dataset 4--Top: Tensor LIC of the principal strain direction overlaid on a tensile test specimen. Bottom: Tensor ellipsoids colored by anisotropy.}
    \label{fig:LIC}
\end{figure}

To visualize the orientation of principal strains, tensor line integral convolution (LIC) is used (Fig.~\ref{fig:LIC}). LIC integrates noise textures along the local major principal strain directions to produce smooth, directionally coherent images~\cite{Hotz2004}. 
Streamline integration is performed with sub-pixel precision using fourth-order Runge–Kutta (RK4).
Aliasing is mitigated through super-sampling, with multiple sub-pixel offsets per pixel.
GPU acceleration via OpenCL enables near real-time rendering of LIC images.

\begin{figure}[!b]
    \centering
    \includegraphics[width=\linewidth]{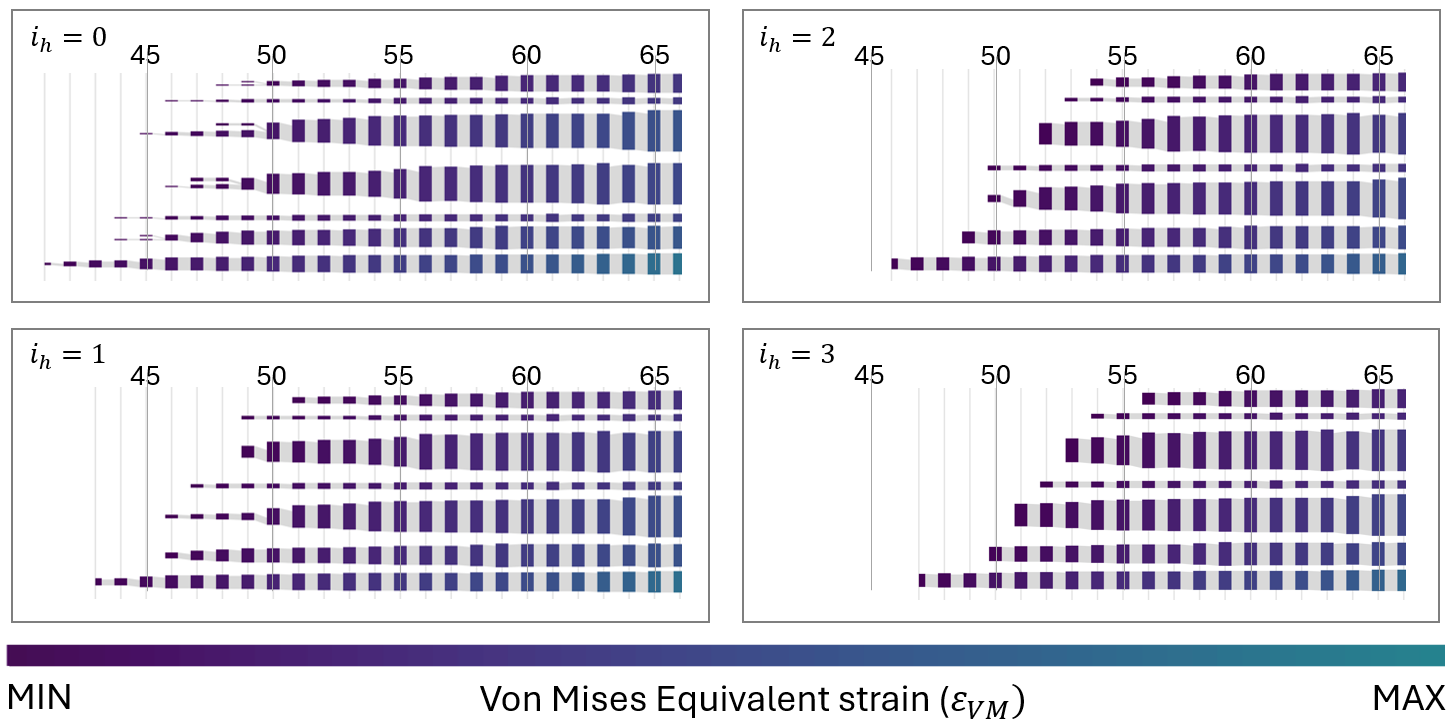}
    \caption{Dataset 4—Comparison across persistence parameters $h$, with differences most during component formation.}
    \label{fig:parameter-h}
\end{figure}
\begin{figure}[!t]
    \centering
    \includegraphics[width=\linewidth]{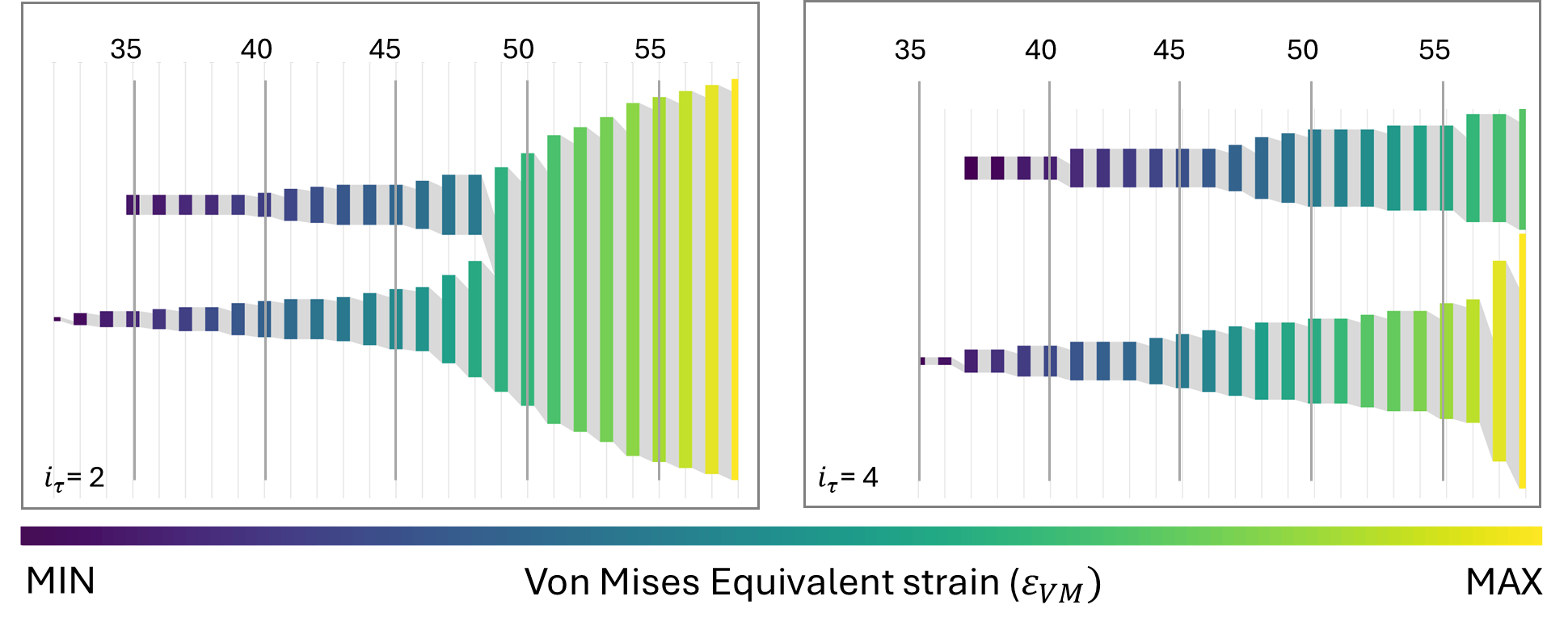}
    \caption{Dataset 3—Comparison across threshold parameters $\tau$, a higher threshold provides more detail, especially close to failure.
    }
    \label{fig:parameter-t}
\end{figure}

\subsection{Topological filtering and parameter selection}

The pipeline employs two key parameters: a superlevel-set threshold $tau$ and a persistence threshold $h$. Their effects are evaluated through parameter sweeps designed to quantify the sensitivity of the method to strain thresholds and noise filtering levels (Fig.~\ref{fig:parameter-h} and~\ref{fig:parameter-t}). The superlevel-set threshold $\tau$ is defined as a fraction of the dataset’s 95th-percentile magnitude $P_95$: $\tau = i_t \frac{P_{95}}{4}, \quad i_t = 1,2,3,4,$. This formulation ensures that the threshold adapts to the scale of each dataset while probing progressively stricter superlevel sets. Similarly, persistence values are parameterized as $ h = i_h \frac{0.125}{3}, \quad i_h = 0,1,2,3,$. Varying $h$ controls the degree of topological noise suppression.


\subsection{Temporal Region Tracking}
\begin{figure}[!htbp]
    \centering
    \includegraphics[width=0.75\linewidth]{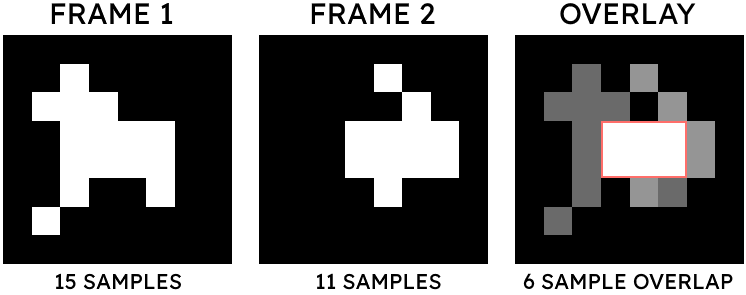}
    \caption{Temporal region tracking using sample overlaps}
    \label{fig:SampleOverlap}
\end{figure}

Superlevel-set regions extracted from thresholded $\epsilon_{VM}$ fields are tracked across frames to capture temporal evolution. For each region in frame $t$, overlaps with regions in frame $t+1$ are computed in the undistorted sample-space (Fig~\ref{fig:SampleOverlap}). Overlaps define temporal relationships, including births, deaths, merges, and splits. Regions with no predecessor in the previous frame are considered births, while regions with no successor in the next frame are considered deaths. Merges occur when multiple regions in frame $t$ overlap a single region in frame $t+1$, and splits occur when a single region in frame $t$ overlaps multiple regions in frame $t+1$. The intersection size, measured in the number of shared samples, is recorded to quantify temporal continuity. 

\vspace*{-.5cm}
\subsection{Sankey Diagram Construction}
\begin{figure}[!b]
    \centering
    \includegraphics[width=1\linewidth]{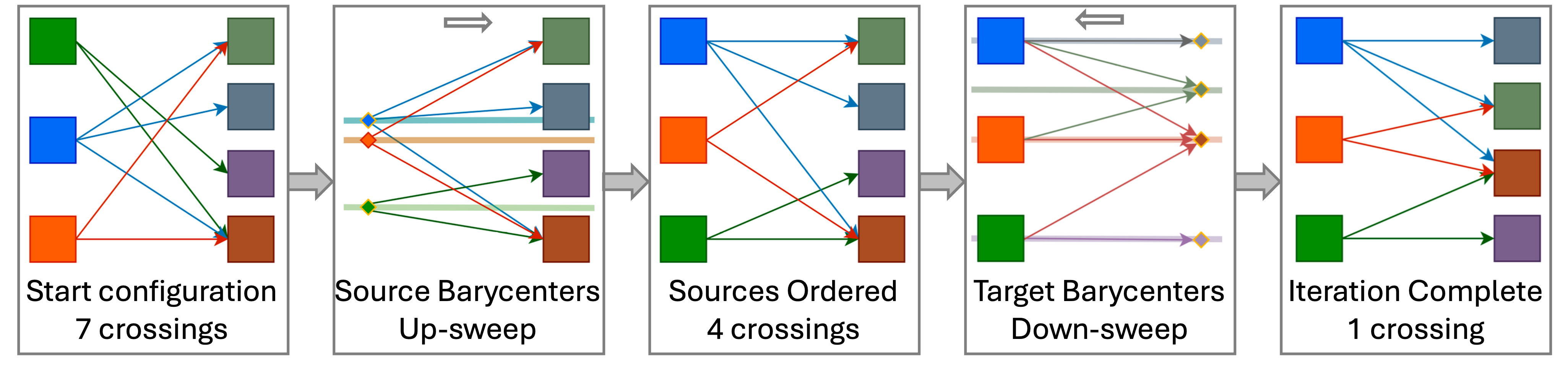}
    \caption{Minimizing link crossings using the barycenter heuristic}
    \label{fig:barycenter}
\end{figure}

The temporal evolution of superlevel-set regions is summarized using a Sankey diagram. Nodes represent connected high von Mises strain regions at frame $t$, and links encode spatial overlap between consecutive frames, capturing contimuation, splitting, and merging.   
Node height corresponds to region area, node color encodes the maximum strain $\epsilon_{VM}$,  and link thickness reflects the number of shared samples, indicating temporal continuity.
Nodes are initially ordered by region ID and scaled to the available space. Link crossings are reduced using the barycenter heuristic~\cite{Makinen2005} (Fig.~\ref{fig:barycenter}), followed by lane-based packing that aligns nodes along dominant flow paths and propagates this alignment backward to ensure temporal consistency.
Finally, nodes in the Sankey diagram can be cross-referenced with their corresponding spatial superlevel-set regions, allowing analysts to relate summarized temporal patterns to the original strain fields.

\begin{figure*}
    \centering
    \includegraphics[width=1\linewidth]{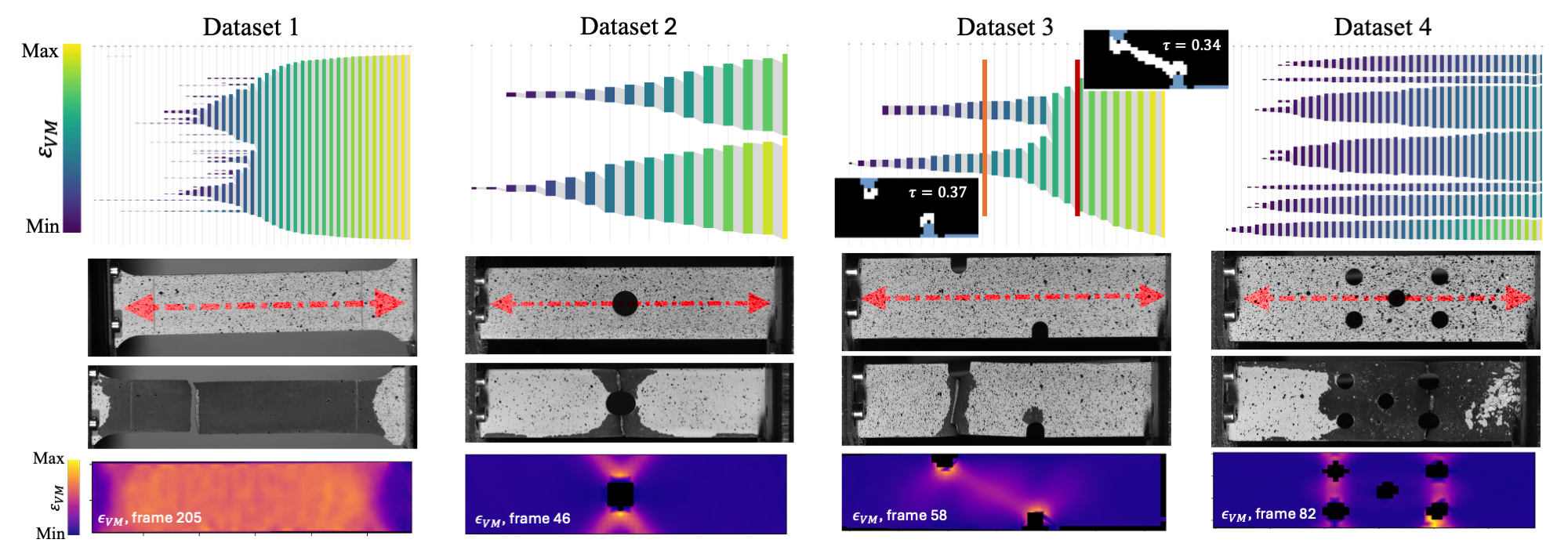}
    \caption{All four experimental specimens are shown before and after failure, together with Sankey diagrams illustrating strain evolution colored by each component’s maximum von Mises equivalent strain ($\epsilon_{VM}$), and a spatial $\epsilon_{VM}$ map shortly before failure. 
    }
    \label{fig:data}
\end{figure*}

\section{Experimental Results}
\label{sec:results}
We evaluate the Sankey-based visualization on four tensile test datasets that demonstrate how geometry affects strain localization and its temporal evolution: D1 (no holes or notches), D2 (one central hole), D3 (two symmetric notches), and D4 (five holes).

\paragraph*{Dataset 1.} In the simplest configuration, D1, the Sankey diagram shows scattered high-strain regions emerging early in loading, which gradually coalesce into two dominant areas that quickly merge into a single large component. This component persists with continuously increasing $\epsilon_{VM}$ reflecting the transition from distributed to global strain localization. 

\paragraph*{Dataset 2.} D2 shows two symmetric high-strain regions around the central hole, emerging nearly simultaneously, indicating independent strain localization (Fig.~\ref{fig:data}, D2). The lower region attains higher $\epsilon_{VM}$  earlier 0.07 to  0.12, (42\% increase), with only a 12\% area increase, while the upper region rises from 0.06 → 0.11 (45\% increase) over a larger area (20\%). Although the first failure cannot be conclusively determined, the stronger localization in the lower region indicates a likely rupture site. Sankey diagrams capture these mirrored, long-lived node-chains reflecting their symmetry, suggesting that concentrated strain rather than total area may indicate regions at higher risk of rupture.

\paragraph*{Dataset 3.} D3 shows antisymmetric strain patterns connecting two notches, forming two "V"-shaped regions that grow and merge over time (Figure~\ref{fig:data} D3, bottom). 
The Sankey diagrams reveal that the bottom node-chain emerges slightly earlier and reaches higher maximum $\epsilon_{VM}$ values than the top chain. However, failure occurs in the region associated with the top node-chain, demonstrating that neither maximum strain nor lifetime alone reliably predicts failure. These observations suggest that while high $\epsilon_{VM}$ indicates the onset of strain localization, it does not necessarily correlate with rupture. Noise from short-lived regions can be mitigated with h-maxima filtering, though the choice of threshold affects interpretation and can obscure the births of larger regions.

\paragraph*{Dataset 4.}  D4 demonstrates the strongest predictive potential for identifying material failure regions. Due to its complex geometry, multiple strain concentrations emerge around holes and specimen edges, producing several node-chains that appear almost simultaneously in the Sankey diagram. While most regions evolve gradually, one region located between the bottom-right notch and the specimen edge exhibits a rapid and highly localized increase in $\epsilon_{VM}$ , reaching the highest strain values among all regions. This localized strain spike precedes the first observed material failure at frame 82, confirming a strong correlation between rapidly intensifying, spatially concentrated strain and rupture initiation.

\section{Conclusion}
This work demonstrated that Sankey-based diagrams can effectively summarize the temporal evolution of strain in tensile test specimens of INCONEL 718, capturing the birth, growth, merging of superlevel-set regions in the von Mises equivalent strain field. 
Across the four datasets, the diagrams revealed persistent node-chains corresponding to strain localization near stress concentrators, with Dataset 4 showing the clearest correspondence between a localized high-strain region and the first failure.
Dominant node-chains remain stable across moderate threshold and persistence settings, indicating robustness to noise; however, manual DIC placement, scale normalization, and limited temporal resolution constrain predictive capability. 
Future work will focus on interactive brushing and linking with spatial maps, improved thresholding strategies, including simultaneous tracking of multiple thresholds in nested diagrams~\cite{Lukasczyk2019} and higher frame-rate acquisition to better resolve strain evolution. Additional directions include application to more complex datasets and quantitative analysis of node-chain metrics to strengthen predictive insight and scientific rigor.

\section{Acknowledgments}
   This work was supported by the Wallenberg AI, Autonomous Systems and Software Program (WASP), funded by the Knut and Alice Wallenberg Foundation. The authors additionally acknowledge support from the Swedish e-Science Research Centre (SeRC) and The Excellence Center at Linköping – Lund in Information Technology (Elliit). The authors also thank Image Systems Motion Analysis for providing the datasets and computational resources.

\bibliographystyle{eg-alpha-doi} 
\bibliography{egbibsample}       

@article{Kratz2013,
	author = {Anderea Kratz and Cornelia Auer and Markus Stommel and Ingrid Hotz},
	doi = {doi.org/10.1111/j.1467-8659.2012.03231.x},
	journal = {Computer Graphics Forum - State of the Art Reports},
	number = {1},
	pages = {49--74},
	title = {Visualization and Analysis of Second-Order Tensors: Moving Beyond the Symmetric Positive-Definite Case},
	volume = {32},
	year = {2013}
}

@article{Yan2021b,
	author = {Lin Yan and Talha Bin Masood and Raghavendra Sridharamurthy and Farhan Rasheed and Vijay Natarajan and Ingrid Hotz and Bei Wang},
	journal = {{Computer Graphics Forum}},
	number = {3},
	pages = {599-633},
	title = {Scalar Field Comparison with Topological Descriptors: Properties and Applications for Scientific Visualization},
	volume = {40},
	year = {2021}
}

@article{Lukasczyk2019,
	author = {Jonas Lukasczyk and Christoph Garth and Gunther H. Weber and Tim Biedert and Ross Maciejewski and Heike Leitte},
	journal = {{IEEE Transactions on Visualization and Computer Graphics (TVCG)}},
	number = {1},
	title = {Dynamic Nested Tracking Graphs},
	volume = {26},
	year = {2019}
}

@inproceedings{Hotz2004,
	author = {Hotz, I and Feng, L and Hagen, H and Hamann, B and Jeremic, B and Joy, K.I},
	publisher = {IEEE Computer Society Press},
    booktitle={IEEE Visualization 2004}, 
	title = {Physically based methods for tensor field visualization},
	year = {2004}}

@inproceedings{Kratz2014,
	author = {Andrea Kratz and Marc Sch\"oneich and Valentin Zobel and Bernhard Burgeth and Gerik Scheuermann and Ingrid Hotz and Markus Stommel},
	booktitle = {Proceedings of Pacific Vis Conference},
	title = {{Tensor Visualization Driven Mechanical Component Design}},
	year = {2014}
}

@article{Schmidt2008,
	author = {Mario Schmidt},
	journal = {Journal of Industrial Ecology},
	title = {The Sankey Diagram in Energy and Material Flow Management, Pert I: Hisotroy},
	year = {2008}
}

@article{Makinen2005,
	author = {Erkki M{\"a}kinen and Harri Siirtola},
	journal = {Informatica},
	title = {The Barycenter Heuristic and the Reorderable Matrix},
	volume = {29},
	year = {2005}
}

@book{Holzapfel2000,
	author = {Holzapfel, GA},
	project = {STAR Tensor II},
	publisher = {Chichester, New York},
	title = {Nonlinear Solid Mechanics},
	volume = {24},
	year = {2000}}

@article{Hergel2021,
	author = {Chiara Hergl and Christian Blecha and Vanessa Kretzschmar and Felix Raith and Fabian G\"unther and Markus Stommel and Jochen Jankowai and Ingrid Hotz and Gerik Scheuermann1},
	doi = {10.1111/cgf.14209},
	journal = {{Computer Graphics Forum}},
	number = {6},
	pages = {135--161},
	title = {Visualization of Tensor Fields in Structural Mechanics and Geology},
	volume = {40},
	year = {2021}}

@book{ImageCorrelationBook,
  Author    = {Michael A. Sutton and Jean-Jos{\'e} Orteu and Hubert W. Schreier},
  Title     = {Image Correlation for Shape, Motion and Deformation Measurements: Basic Concepts, Theory and Applications},
  Publisher = {Springer Science+Business Media LLC},
  Year      = {2009},
  ISBN      = {978-0-387-78746-6},
  DOI       = {10.1007/978-0-387-78747-3}
}

@book{MechanicalPhysicsBook,
  Author    = {Ferdinand P. Beer and E. Russell Johnston, Jr},
  Title     = {Mechanics of Materials: Second Edition in SI Units},
  Publisher = {McGraw-Hill Book Company Europe},
  Edition   = {Second Edition},
  Year      = {1992},
  ISBN      = {0-07-112939-1}
}

@misc{DICpracticesGuide,
  Author      = {International Digital Image Correlation Society},
  Editor      = {Jones, E.M.C. and Iadicola, M.A.},
  Title       = {A Good Practices Guide for Digital Image Correlation},
  Year        = {2018},
  Type        = {Technical Report},
  DOI         = {10.32720/idics/gpg.ed1},
  URL         = {https://doi.org/10.32720/idics/gpg.ed1}
}


\end{document}